# Transverse orbital angular momentum of spatiotemporal optical vortices: setting the record straight

**N. TRIPATHI, S.W. HANCOCK, AND H.M. MILCHBERG***

*Institute for Research in Electronics and Applied Physics and Dept. of Physics, University of Maryland, College Park, Maryland 20742, USA*
*milch@umd.edu

**Abstract:** The nature of the transverse orbital angular momentum (tOAM) associated with spatiotemporal optical vortex (STOV) pulses has been the subject of recent debate. We demonstrate that the approaches to tOAM presented in several recent papers are incorrect and lead to unphysical results, including erroneous claims of zero total tOAM for a freely propagating STOV pulse. We emphasize the importance of calculating the OAM of any extended physical object at a common instant of time, and reemphasize the special status of the centre of energy as a reference point for all OAM calculations. The theory presented in [Phys. Rev. Lett. **127**, 193901 (2021)] is the only correct classical field-based framework that both agrees with experiments and provides a self-consistent understanding of transverse OAM in spatiotemporal light fields.

## 1. Introduction

Optical beams with a phase singularity in the plane orthogonal to the propagation direction possess longitudinal orbital angular momentum (OAM) directed along or opposed to propagation. A well-known example is Laguerre-Gaussian beams [1] with an azimuthal phase circulation $e^{im\varphi}$, where there are integer $m$ units of OAM per photon. Such beams can be monochromatic or polychromatic (pulsed). Less well known, but with rapidly increasing interest and activity, are optical pulses with transverse OAM (tOAM), where the phase circulation is in a spatiotemporal plane. Such electromagnetic structures, dubbed “STOVs” (spatiotemporal optical vortices) because their phase winding is described in spacetime coordinates, were first observed as an emergent effect generated by extreme spatiotemporal phase shear in femtosecond filamentation in air [2]. Since STOVs are carried by light pulses of finite duration, they are polychromatic [3]. STOVs are naturally emergent in any arrested nonlinear self-focusing process such as filamentation in air [2] and relativistic self-guiding in plasmas [4], and once generated, they play a dominant role in controlling intrapulse energy flows [2,4]. STOV dynamics and their topological constraints can be viewed as providing a unifying principle explaining complex phenomena such as pulse splitting in self-focused propagation [4,5].

The realization that STOVs were formed by spatiotemporal phase shear [2] led us to design a $4f$ pulse shaper that linearly imposes spatiospectral phase shear and then returns the pulse to the spatiotemporal domain [6-8]. The resulting free-space-propagating STOV pulses [8] were measured from the near to far field by a single-shot method well-suited to pulses with spacetime singularities [9]. Later work used a similar pulse shaper and a multi-shot scanning technique to measure STOVs in the far field only [10]. A more recent single shot technique, resolution-limited by the pulse bandwidth, used spatially resolved spectral interferometry to characterize STOVs [11]. In further work, tOAM conservation in second-harmonic generation [12-15] has been demonstrated, alternative methods for STOV generation have been proposed [16-18], and applications such as high-harmonic generation [19-21], laser wakefield acceleration [22,23], and acoustics [24,25] have been explored. Most recently, experiments and simulations have revealed how matter can spatiotemporally torque light [26,27] and vice versa [28], with the latter leading to the excitation of STOV polaritons in dispersive media.

With research activity in structured light continuing to increase [29], it is timely to assess several recent theoretical approaches to tOAM: Hancock *et al*. [30], Bliokh [31,32], and later work by Porras [33,34]. The goal of this paper is to briefly review these theories, discuss significant errors in [31-34], and re-introduce clarity to the discussion of transverse orbital angular momentum.

## 2. Transverse OAM: basic considerations

In parallel with the experimental developments in STOVs, there has been a debate about the nature of transverse OAM and how to calculate it. To set the stage for the discussion to follow, we briefly review some basic considerations, starting with the definition (in Gaussian units) of total electromagnetic OAM [35]

$$\mathbf{L} = (4\pi c)^{-1} \int d^3\mathbf{r}\, \mathbf{r} \times (\mathbf{E} \times \mathbf{H}), \tag{1}$$

where $\mathbf{E} = \mathbf{E}(\mathbf{r}, t)$ and $\mathbf{H} = \mathbf{H}(\mathbf{r}, t)$ are the electric and magnetic fields. While perhaps obvious, it is important to note that the fields are integrated over all space at the common lab time $t$; $\mathbf{L}$ is meant to be an instantaneous quantity at time $t$. In general, calculating *any* aggregate physical quantity for a composite object demands summation or integration over all the constituent contributions *at the same instant of time*. For example, if the total linear momentum of a cloud of gas was calculated by summing molecular contributions from different times, one would need to specify those times; in any case, such a quantity would generally not be conserved, severely limiting its utility. As we will show, it is precisely this misapprehension about the correct way to sum OAM contributions of a composite object, here a light pulse, that underlies the errors of [33,34].

Starting with Eq. (1), the per-photon *intrinsic* (origin-independent) *j*-component of OAM at laboratory time $t$ can be calculated from the equivalent approaches [26] of Eq. (2a) or (2b):

$$\langle L_j \rangle = 2k_0 U^{-1} \int d^3\boldsymbol{\mathcal{R}}\, [(\mathbf{r} - \mathbf{r}_c) \times (\mathbf{E} \times \mathbf{H})]_j \tag{2a}$$

$$\langle L_j \rangle = u^{-1} \langle E | L_j | E \rangle \quad . \tag{2b}$$

Here $\mathbf{r}_c = U^{-1} \int d^3\boldsymbol{\mathcal{R}}\, \mathbf{r}(|\mathbf{E}|^2 + |\mathbf{H}|^2)$ is the energy centroid (or centre of energy), $\langle\ \rangle$ denotes expectation value, $U = \int d^3\boldsymbol{\mathcal{R}}\, (|\mathbf{E}|^2 + |\mathbf{H}|^2)$ is proportional to pulse energy, $\langle E | L_j | E \rangle = \int d^3\boldsymbol{\mathcal{R}}\, E^* L_j E$, $L_j$ is the *j*-component OAM operator (to be discussed shortly), $u = \langle E | E \rangle = \int d^3\boldsymbol{\mathcal{R}}\, |E|^2$, and $d^3\boldsymbol{\mathcal{R}}$ is a volume element in laboratory frame coordinates, which takes on specific forms depending on the type of OAM calculated. The origin used in the integral of Eq. (2b) is the centre of energy.

For *longitudinal* OAM of a monochromatic beam or polychromatic pulse propagating along $z$, $L_j = L_z = (\mathbf{r} \times \hat{\mathbf{p}})_z = -i(x\,\partial/\partial y - y\,\partial/\partial x)$ is the longitudinal OAM operator in Eq. (2b), using $\hat{\mathbf{p}} = -i\boldsymbol{\nabla}$ as the linear momentum operator. This form of $L_z$ guarantees that OAM results from Eq. (2b) and Eq. (2a) agree. Physically, this operator admits energy density circulation in the $xy$ plane, transverse to propagation. For monochromatic beams, $d^3\boldsymbol{\mathcal{R}} \rightarrow d^2\mathbf{r}_\perp$, an area element transverse to propagation, while for pulses one can use in Eqs. (2a) and (2b) either of the lab frame-based elements $d^3\boldsymbol{\mathcal{R}} = d^3\mathbf{r}$ or $d^3\boldsymbol{\mathcal{R}} = d^2\mathbf{r}_\perp d\xi$, where $\xi = v_g t - z$ is a space coordinate local to the pulse (based on lab frame coordinates $z$ and $t$), $v_g = (\partial k/\partial\,\omega)^{-1}_{\omega_0}$ is the pulse group velocity, $\omega_0$ is the pulse central frequency, and $k_0 = k(\omega_0)$ is the central wavenumber. Applying either of Eqs. (2a) or (2b) to a Laguerre-Gaussian mode of integer azimuthal winding number $m$ gives $m$ units of OAM per photon.

*Transverse* OAM embedded in a propagating pulse (as distinct from stationary transverse OAM [36,37]) is associated with spatiotemporal optical vortices (STOVs), an essentially polychromatic structure. STOV-carrying electromagnetic pulses can be constructed as linear combinations of spatiotemporal Hermite-Gaussian solutions of the paraxial spatiotemporal wave equation [30],

$$2ik_0\,\partial\mathbf{A}/\partial z = (-\nabla_\perp^2 + \beta_2\,\partial^2/\partial\xi^2)\mathbf{A} = H\mathbf{A} \tag{3}$$

describing linear propagation along $z$ in a dispersive medium. Here $\mathbf{A} = \mathbf{A}(\mathbf{r}_\perp, \xi; z)$ is the slowly varying envelope of the vector potential ($\mathbf{A} = -i(c/\omega_0)\mathbf{E}$), $H = -\nabla_\perp^2 + \beta_2\,\partial^2/\partial\xi^2$ is the propagation operator, and $\beta_2 = v_g^2 k_0 k_0''$ is the dimensionless group velocity dispersion of the medium, with $k_0'' = (\partial^2 k/\partial\,\omega^2)_{\omega_0}$. The separation of $z$ by a colon in $\mathbf{A}(\mathbf{r}_\perp, \xi; z)$ denotes its role as a time-like running parameter. Reference [30] showed from first principles––using conservation of energy density flux and tOAM––that the $y$-component of the transverse OAM operator is $L_y = (\mathbf{r}' \times \widehat{\mathbf{p}}_{st})_y = -i(\xi\,\partial/\partial x + \beta_2 x\,\partial/\partial\xi)$, where $\mathbf{r}' = (\mathbf{r}_\perp, \xi) = (x, y, \xi)$ and $\widehat{\mathbf{p}}_{st} = -i(\boldsymbol{\nabla}_\perp - \widehat{\boldsymbol{\xi}}\,\beta_2\,\partial/\partial\xi)$ is the spacetime linear momentum operator ($\widehat{\boldsymbol{\xi}}$ is a unit vector along $\xi$). This form of $L_y$ ensures agreement between the results of Eqs. (2a) and (2b), where one can use either $d^3\boldsymbol{\mathcal{R}} = d^3\mathbf{r}$ or $d^3\boldsymbol{\mathcal{R}} = d^2\mathbf{r}_\perp d\xi$. Importantly, because $d/dz\,\langle L_y\rangle = i(2k_0)^{-1}\langle[H, L_y]\rangle = 0$, where $[H, L_y]$ is the commutator of the two operators, $L_y$ is conserved with propagation along $z$.

Near the beam waist (for $z/z_{0x} \ll 1$ and $z/z_{0y} \ll 1$), a STOV solution to Eq. (3) for a dilute medium or vacuum ($\beta_2 = 0$) is

$$\mathrm{A}(x, y, \xi; z) \;= \mathrm{A}_0\left(\frac{\xi}{w_{0\xi}} + \mathrm{sgn}(l)\frac{ix}{w_{0x}}\right)^{|l|} e^{-(x^2/w_{0x}^2 + y^2/w_{0y}^2)} e^{-\xi^2/w_{0\xi}^2}\, e^{-ik_0\xi}\,, \tag{4}$$

where $w_{0x}$ and $w_{0\xi}$ are spacelike and timelike pulse widths, $z_{0x} = \frac{1}{2}k_0 w_{0x}^2$ and $z_{0y} = \frac{1}{2}k_0 w_{0y}^2$ are Rayleigh ranges associated with the transverse beam widths (here we assume $w_{0x} \ll w_{0y}$ and neglect the $y$-dependence), and $l = 0, \pm1, \pm2, \ldots$ is the spatiotemporal vortex winding number. In vacuum or in a negligibly dispersive material, $L_y = -i\xi\,\partial/\partial x$, which admits only transverse (along $x$) energy density circulation. For the STOV pulse of Eq. (4) with $\alpha = w_{0\xi}/w_{0x}$, both Eq. (2a) and (2b) give $\langle L_y\rangle = \frac{1}{2}l\alpha$ [26,30]. The *half-integer* dependence originates from restriction of vortical energy density circulation to $\pm x$ only. Circulation along $\xi$ in the moving frame in vacuum would violate special relativity [26]. For the case of nonzero $\beta_2$, dispersion *will* mediate energy transport along $\xi$ and the tOAM per photon becomes $\langle L_y\rangle = \frac{1}{2}l(\alpha - \beta_2/\alpha)$.

In our experiments [8,26,27,30], measurements of STOVs were resolved in $(x, \xi)$ coordinates by our single shot diagnostic TG-SSSI (transient grating single shot supercontinuum spectral interferometry) [9]. This diagnostic imprints––at a common lab time $t$––the STOV field's entire amplitude and phase spatial dependence (transversely resolved along $x$ and longitudinally resolved along $\xi$) on a longer, co-propagating probe pulse. For definiteness, one can consider $\xi = 0$ to be the location of the pulse centre of energy; for a symmetric STOV pulse this is co-located with the vortex singularity, which propagates a distance $z = v_g t$. Then $z$ $(= v_g t)$ can be considered as a running parameter "label" applying to the whole moving window. As the extracted spatial field distribution in $(x, \xi)$ *applies at a common instant of time*, it enables the correct calculation of tOAM via Eqs. (2a) or (2b). We will return to this discussion later in the context of refs. [33,34].

## 3. Assessment of some recent theories of transverse OAM

In refs. [31,32], the operator for intrinsic tOAM is asserted to have the same form as for longitudinal OAM, namely $\pounds_y = -i(\xi\,\partial/\partial x - x\,\partial/\partial\xi)$, to which we assign a different symbol to keep it distinct from $L_y$. This operator admits energy density circulation along both $x$ and $\xi$ axes in vacuum and thus incorrectly predicts *integer* transverse OAM. Furthermore, $\langle\pounds_y\rangle$ is not conserved in propagation, nor does the computed tOAM using $\pounds_y$ agree with the results of spatiotemporal torque experiments [26]. In additional work [32] intended to support the *integer* theory of [31], it was argued that the correct choice of origin in determining intrinsic tOAM is not the centre of energy, but the "photon centroid" [32]. However, this choice of origin does not fix the non-conservation of $\langle \pounds_y \rangle$ and still results in an incorrect value for the intrinsic tOAM [26]. A complete discussion of errors from using $\pounds_y$ and a non-energy-centre origin is given in the Appendices of [26] and need not be reproduced here. In Appendix A of this paper, we address the use of the photon centroid as defined in [32].

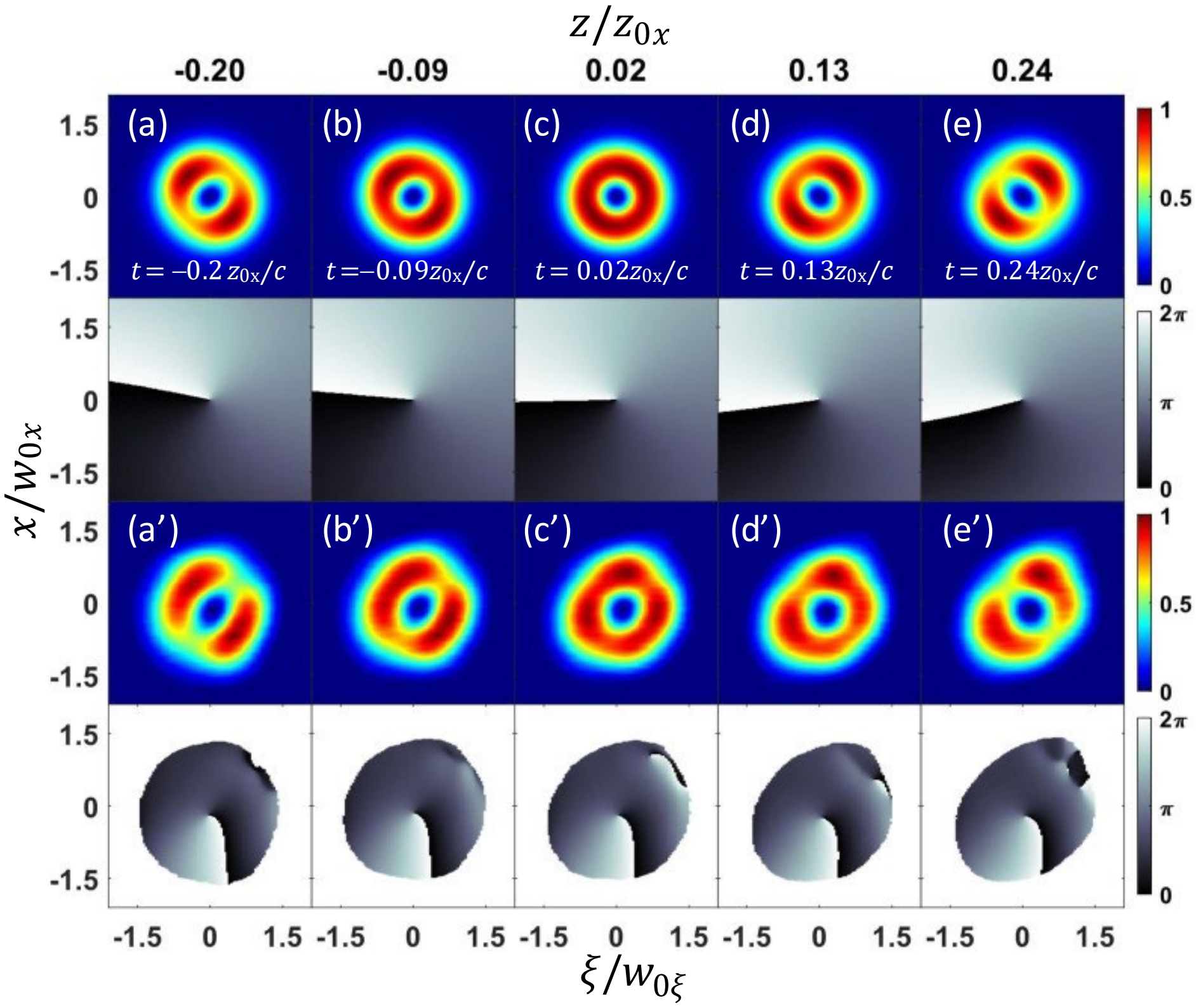

**Fig. 1.** Comparison of analytic and experimental STOV pulse propagation for $l = 1$, $\alpha = w_{0\xi}/w_{0x} = 0.24$, with $z_{0x} = k_0 w_{0x}^2/2 = 4.5$ cm. The top two rows, (a)-(e), show the analytic solution [30] plots for the spatiotemporal intensity (top) and phase (below). The panels show propagation locations from $z/z_{0x} = -0.20$ to $z/z_{0x} = 0.24$ (where $z = v_g t = ct$ for $t = -0.20\, z_{0x}/c$ to $t = 0.24\, z_{0x}/c$). Within each frame the pulse propagates from right to left. The bottom two rows, (a′) - (e′), show experimental intensity and phase plots recorded using TG-SSSI [9].

Several years after the *integer* vs. *half-integer* debate began, an approach based on angular momentum flux density, introduced in [38] for longitudinal OAM, was applied to STOVs [33]. That result calculated an *extrinsic* tOAM of equal amplitude but opposite sign to the intrinsic tOAM, resulting in *zero total tOAM*. The author then generally concluded that a STOV pulse would not be able to convey tOAM to matter and rotate it—basing a claim with real physical consequences on an arbitrary choice of origin for extrinsic OAM [33]. Our recent work [28] demonstrating tOAM transfer from a STOV pulse to particles disproves this claim. The author of [33] has continued to claim that STOVs have zero total tOAM, more recently publishing papers entitled "*Clarification of the transverse orbital angular momentum of spatiotemporal optical vortices*" [34] and "*Closing the debate on the transverse orbital angular momentum of* *spatiotemporal optical vortices*" [39], purporting to both resolve the debate and reveal errors in our work [26,30]. Some of these arguments have been further propagated in a recent review article [40]. Several claims in [33,34] are new and we address them below.

We replot in Fig. 1 spatiotemporal intensity and phase results from analytic solutions to Eq. (3) and experiment [26,30] for free space propagation of an $l = 1$ STOV with $\alpha = w_{0\xi}/w_{0x} = 0.24$. The plots are resolved in $(x, \xi)$, with the common lab frame time $t$ within each panel labeled in the top row. The pulse propagates right to left within each panel. The theory plots (a)-(e) are constructed from spatiotemporal Hermite-Gaussian solutions to Eq. (3), which, near $z = 0$, approach the form of $A(x, y, \xi; z)$ in Eq. (4). The agreement between experiment (Fig. 1(a′)-(e′)) and theory (Fig. 1(a)-(e)) validates not only our theory, but also our use of $(x, \xi)$ coordinates, which are (1) native to our TG-SSSI diagnostic, (2) natural to the spatiotemporal wave equation [30] (Eq. (3)), and (3) the appropriate coordinates to use in computation of tOAM using Eqs. (2a) and (2b).

To compare the results of our theory and the calculations by Porras in [33,34,39], we plot in Fig. 2(a)-(e) computed intensity profiles of a STOV pulse using the parameters of [34] ($l = 1$, $w_{0x} = 10\ \mu m$, $w_{0\xi} = 75\ \mu m$, $\alpha = 7.5$) for times running from $t = -2t_0$ to $2t_0$ in steps of $t_0 = z_R/16c$. The profiles (plotted as 80% iso-intensity surfaces as in [34]) are computed from the spacetime Hermite-Gaussian modal solutions to the spatiotemporal paraxial wave equation (Eq. (3)) [30], as done in Fig. 1(a)-(e). The pulse propagates right to left in each panel, where the vertical red line marks the beam waist location $z = 0$, which is fixed in the lab frame. Each of the profiles in Fig. 2(a)-(e) is a spatial snapshot in $(x, \xi)$ to which a common time $t$ applies. The associated field profiles are therefore appropriate for insertion into either Eq. (2a) or (2b) for calculation of tOAM per photon.

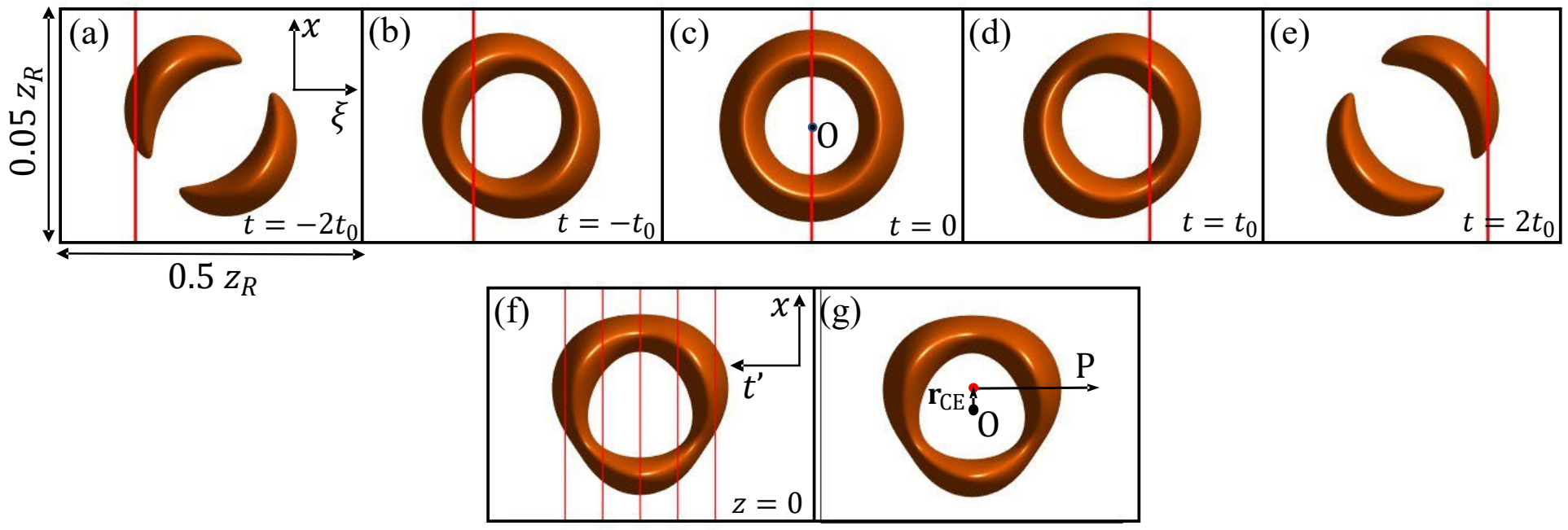

**Fig. 2.** (a)-(e): Sequence of frames (80% maximum iso-intensity surfaces) of $l = 1$ STOV pulse propagating through its beam waist at $z = 0$ in the lab frame, depicted as the vertical red line. The frames are at times $t = -2t_0$, $-t_0$, 0, $t_0$, and $2t_0$ (with $t_0 = z_R/16c$), and are resolved in $(x, \xi)$, the natural coordinates for our TG-SSSI measurements [9], our tOAM theory [30], and for calculations of tOAM using Eq. 2(a) or 2(b) (see text). The STOV parameters of [34] are used here ($l = 1$, $w_{0x} = 10\ \mu m$, $w_{0\xi} = 75\ \mu m$, $\alpha = 7.5$). (f) Artificial lopsided intensity profile constructed from sampling the propagating electric field at $z = 0$ at successive times spaced by $\Delta t = z_R/1000c$, and plotting the squared magnitude of the composite field. Each of the five vertical red lines in (f) correspond to the vertical red lines in (a)-(e) and those particular sampling times. (g) Centre of energy $\mathbf{r}_{CE}$ of artificial lopsided structure of (f) with respect to true centre of energy (O) and linear momentum **P** drawn as in Fig. 2(b) in Ref. [34]. The $(x, t')$ axes applying to (f) and (g) highlight that these lopsided intensity profiles are constructed from a range of times.

In Fig. 2(f), we have constructed the exact same lopsided structure that appears in Fig. 2(b) of ref. [34]. The claim of Porras in [34] is that what we measure and calculate as a symmetric STOV profile near the beam waist in Figs. 1(c,c') and 2(c) (along with symmetric evolution away from the waist) is in reality the structure in Fig. 2(f), with its centre of energy $\mathbf{r}_{CE}$ displaced from the z axis as shown in Fig. 2(g), along with the same momentum vector drawn in Porras's figure. This structure purportedly supplies a non-zero extrinsic tOAM, $\mathbf{r}_{\mathrm{CE}} \times \mathbf{P}$, that cancels the intrinsic tOAM, yielding a total tOAM of zero. Porras [34] states that "*STOVs that are elliptical* (symmetric) *in space-time at a transversal plane do not carry transverse OAM about the origin O because the extrinsic and the intrinsic transverse OAM are opposite*", while the abstract in [33] announces that this result "*may preclude applications such as setting particles into rotation*".

So how did we construct Fig. 2(f) to exactly get Porras's lopsided STOV result in [34]? In Fig. 2(a)-(e), the pulse passes though $z = 0$ (vertical red line) at the sequence of times labeled in the frames. Figure 2(f) is an artificial intensity profile constructed from sampling the electric field at $z = 0$ at successive times spaced by $\Delta t = z_R/1000c$, and then plotting the squared magnitude of the composite field. The vertical lines in Fig. 2(f) correspond to the particular sampling times in Fig. 2(a)-(e). This is effectively the same procedure used by Porras to construct Fig. 2(b) in [34]. That is, Porras's lopsided structure *is constructed from field contributions from different times*. This structure is therefore completely unsuitable for a calculation of tOAM, and any conclusions drawn from such a calculation, such as extrinsic tOAM cancelation of intrinsic tOAM [33,34], are incorrect. Appendix B explains the errors in [33,34] as originating from an incorrect application of the momentum flux approach [38], while Appendix C presents a simple model comparing the correct and incorrect approaches to calculating tOAM.

## 4. Conclusions

In summary, the theories of transverse OAM presented in [31-34] are incorrect, adopting unphysical OAM operators, exhibiting non-conservation of tOAM, using unphysical choices of origin, deriving spurious extrinsic tOAM, and constructing unphysical lopsided STOVs. These theories suffer from conceptual and mathematical errors plus misleading and unphysical interpretations, such as that particles cannot be rotated by STOV pulses, and even leading their authors to claim that neither extrinsic nor intrinsic OAM need be conserved for freely propagating pulses [34,41].

The theory presented in [30] is the only correct classical field-based framework that both agrees with experiments and simulations [26-28,30] and provides a self-consistent understanding of transverse OAM in spatiotemporal light fields. For a symmetric STOV pulse ($\alpha = 1$) with winding $l = 1$, both the canonical approach (Eq. (2a)) and the operator approach (Eq. (2b)) give an intrinsic tOAM per photon of ½, and an extrinsic tOAM of zero when the origin is the centre of energy. These conclusions are not only relevant for optical vortices but also extend to other areas, including spatiotemporal acoustic vortices [24,25] and quantum mechanical wavefunctions [32].

## Appendix A: Is the photon centroid a useful quantity for calculating OAM?

In [26], we showed that extrinsic tOAM defined using the "photon centroid" [32] is not conserved. In [32], Bliokh asserts that the *total* tOAM per photon for a $l = 1$, $\alpha = 1$ STOV pulse is $\langle £_y \rangle = \langle £_y^{(i)} \rangle + \langle £_y^{(e)} \rangle = 1/2$, where he computes the intrinsic tOAM $\langle £_y^{(i)} \rangle = 1$ and the extrinsic tOAM as $\langle £_y^{(e)} \rangle = (\mathbf{r}_{\mathrm{PC}} \times \mathbf{p})_y = -1/2$, where $\mathbf{p}$ is the pulse linear momentum per photon and the photon centroid is defined [32] as

$$\mathbf{r}_{\mathrm{PC}}(z) = \frac{\int d^3\mathbf{k}\, \omega^{-1} \widetilde{\mathrm{A}}^* i \nabla_k \widetilde{\mathrm{A}}}{\int d^3\mathbf{k}\, \omega^{-1} \left|\widetilde{\mathrm{A}}\right|^2} \quad . \tag{5}$$

Here $\mathbf{k} = (k_x, k_y, k_\xi)$, $\omega = \omega(\mathbf{k}) = c(k_x^2 + k_y^2 + k_\xi^2)^{1/2}$, and $\widetilde{\mathrm{A}}(\mathbf{k}; z)$ is the Fourier transform of a solution $\mathrm{A}(x, y, \xi; z)$ to the electromagnetic wave equation. Above, we use the symbol $\pounds_y$ to keep it distinct from $L_y$. For later use in this appendix, we can also write $\langle \pounds_y^{(e)} \rangle_{pulse} = \mathbf{r}_{\rm PC} \times \mathbf{P}$ for the whole pulse, as done in [32], where $\mathbf{P}$ is the pulse total linear momentum. In [32], Bliokh claims that what we compute as half-integer intrinsic tOAM in [30] is actually the total (intrinsic plus extrinsic) tOAM, in agreement with his result $\langle \pounds_y \rangle = 1/2$. The claims of [32] rest on the invariance of $\langle \pounds_y^{(e)} \rangle$ with propagation, a question we again examine, this time directly using the definition (Eq. (5)) provided for $\mathbf{r}_{\rm PC}$ in [32].

If $\mathbf{N}$ and $D$ are the numerator and denominator of Eq. (5), then $d\mathbf{r}_{\rm PC}/dz = D^{-1}\, d\mathbf{N}/dz - D^{-2}(dD/dz)\mathbf{N}$. Then $dD/dz = \int d^3\mathbf{k}\, \omega^{-1}\left[\left(\partial \widetilde{\mathrm{A}}^*/\partial z\right)\widetilde{\mathrm{A}} + \widetilde{\mathrm{A}}^*\left(\partial \widetilde{\mathrm{A}}/\partial z\right)\right] = 0$, where we have substituted $\partial \widetilde{\mathrm{A}}/\partial z = (2ik_0)^{-1}\,(k_\perp^2 - \beta_2 k_\xi^2)\widetilde{\mathrm{A}}$, obtained from the Fourier transform in $x, y$, and $\xi$ of the spatiotemporal propagation equation (Eq. (3)). Similarly, it is straightforward to show that $d\mathbf{N}/dz = \int d^3\mathbf{k}\, \omega^{-1}\left[\left(\partial \widetilde{\mathrm{A}}^*/\partial z\right) i\nabla_k \widetilde{\mathrm{A}} + \widetilde{\mathrm{A}}^* i\nabla_k\left(\partial \widetilde{\mathrm{A}}/\partial z\right)\right] = (2k_0)^{-1}\int d^3\mathbf{k}\, \omega^{-1}\left|\widetilde{\mathrm{A}}\right|^2 \nabla_k k_\perp^2$, where in the last step we take $\beta_2 = 0$ (vacuum propagation). This result leads to

$$\frac{d\mathbf{r}_{\rm PC}}{dz} = k_0^{-1}\frac{\int d^3\mathbf{k}\, \omega^{-1}\left|\widetilde{\mathrm{A}}\right|^2 \mathbf{k}_\perp}{\int d^3\mathbf{k}\, \omega^{-1}\left|\widetilde{\mathrm{A}}\right|^2} \qquad (6)$$

for a general origin in the moving frame. Alternatively, one can calculate the centre of energy $\mathbf{r}_{\rm CE}(z) = \int d^3\mathbf{r}\, \mathrm{A}^*\mathbf{r}\mathrm{A} / \int d^3\mathbf{r}\, |\mathrm{A}|^2$, which can also be computed in $\mathbf{k}$-space as

$$\mathbf{r}_{\rm CE}(z) = \frac{\int d^3\mathbf{k}\, \widetilde{\mathrm{A}}^* i\nabla_k \widetilde{\mathrm{A}}}{\int d^3\mathbf{k}\, \left|\widetilde{\mathrm{A}}\right|^2} \quad . \qquad (7)$$

Calculating $d\mathbf{r}_{\rm CE}/dz$ using the same steps that led to Eq. (6) gives

$$\frac{d\mathbf{r}_{\rm CE}}{dz} = k_0^{-1}\frac{\int d^3\mathbf{k}\, \left|\widetilde{\mathrm{A}}\right|^2 \mathbf{k}_\perp}{\int d^3\mathbf{k}\, \left|\widetilde{\mathrm{A}}\right|^2} \quad , \qquad (8)$$

also for a general origin in the moving frame. The numerator of Eq. (8) is proportional to $\mathbf{P}_\perp$, the total perpendicular linear momentum of the pulse, so that $d\mathbf{r}_{\rm CE}/dz \propto \mathbf{P}_\perp$ in the moving frame. However, in general $d\mathbf{r}_{\rm PC}/dz$ is *not* proportional to $\mathbf{P}_\perp$ because of the $\omega^{-1}$ factor in the numerator integral in Eq. (6). (Note that if $\mathbf{r}_{\rm CE}$ is taken as the local origin, $d\mathbf{r}_{\rm CE}/dz = 0$ and $\mathbf{P}_\perp = 0$.)

Then, in the moving frame, we have $d\langle \pounds^{(e)} \rangle_{pulse}/dz = d\mathbf{r}_{\rm PC}/dz \times \mathbf{P} + \mathbf{r}_{\rm PC} \times d\mathbf{P}/dz = d\mathbf{r}_{\rm PC}/dz \times \mathbf{P}_\perp \neq 0$ in general, where we used the facts that $\mathbf{P}_\parallel = 0$ in the moving frame and $d\mathbf{P}/dz = 0$ from propagation invariance of $\mathbf{P}$. By contrast, $d\langle \mathbf{L}^{(e)} \rangle_{pulse}/dz = d\mathbf{r}_{\rm CE}/dz \times \mathbf{P} + \mathbf{r}_{\rm CE} \times d\mathbf{P}/dz = d\mathbf{r}_{\rm CE}/dz \times \mathbf{P}_\perp = 0$ *irrespective of the choice of local origin*. That is, while the centre-of-energy-based extrinsic tOAM, $\langle \mathbf{L}^{(e)} \rangle_{pulse}$, is always conserved, the photon-centroid-based extrinsic OAM, $\langle \pounds^{(e)} \rangle_{pulse}$, is not.

In the *lab frame*, we write $\mathbf{r}_{\rm PC} = x_{\rm PC}\hat{\mathbf{x}} + z_{\rm PC}\hat{\mathbf{z}}$, $\mathbf{r}_{\rm CE} = x_{\rm CE}\hat{\mathbf{x}} + z_{\rm CE}\hat{\mathbf{z}}$, and take $\mathbf{P} = \mathrm{P}\hat{\mathbf{z}}$, neglecting y-dependence by considering a line STOV $\mathrm{A}(x, \xi; z)$. Then the photon centroid definition of extrinsic OAM [32] gives

$$d\langle \pounds_{lab}^{(e)} \rangle_{pulse}/dz = -\hat{\mathbf{y}}\ \mathrm{P}\ dx_{\rm PC}/dz \ , \qquad (9a)$$

and for the centre of energy,

$$d\langle \mathbf{L}_{lab}^{(e)} \rangle_{pulse}/dz = -\hat{\mathbf{y}}\ \mathrm{P}\ dx_{\rm CE}/dz = 0 \qquad (9b)$$

where we have used $d\mathbf{P}/dz = 0$ (from propagation invariance of $\mathbf{P}$) and $dx_{\mathrm{CE}}/dz = 0$ ($d\mathbf{r}_{\mathrm{CE}}/dz \propto \mathbf{P}$ in the lab frame [26]). Thus, while $\langle \mathbf{L}_{lab}^{(e)} \rangle_{pulse}$ is conserved with propagation, $\langle £_{lab}^{(e)} \rangle_{pulse}$ is not.

To summarize, the photon-centroid-based extrinsic tOAM $\langle £^{(e)} \rangle$ per photon (or equivalently, $\langle £^{(e)} \rangle_{pulse}$ for the whole pulse) varies with propagation, rendering $\langle £_y \rangle = \langle £_y^{(i)} \rangle + \langle £_y^{(e)} \rangle$ time dependent unless variations in $\langle £_y^{(i)} \rangle$ exactly cancel those in $\langle £_y^{(e)} \rangle$. Because both total and intrinsic tOAM should be conserved to be useful quantities, we conclude that $\langle £_y \rangle$ and $\langle £_y^{(i)} \rangle$, like $\langle £_y^{(e)} \rangle$, are not useful quantities. We reemphasize that: (1) Use of the photon centroid in calculations of tOAM is a mathematical and conceptual error; only the energy centroid (or energy centre) as a reference origin sets $\mathbf{L}^{(e)} = 0$. (2) Independent of (1), $£_y^{(i)}$ is an incorrect operator for intrinsic tOAM: it is not conserved with propagation as shown in [26] and here, and it is in conflict with canonical and operator calculations of tOAM (Eqs. 2(a) and 2(b)), both of which yield propagation invariant $\langle L_y^{(i)} \rangle = 1/2$ for a symmetric STOV.

### Appendix B: Errors in refs. [33,34]

In Ref. [38], using a flux approach, Barnett calculated the rate of change in longitudinal OAM passing through a surface,

$$\partial J_i/\partial t = -\int_S M_{ji}\, dS_j, \tag{10}$$

where $J_i$ is the $i^{th}$ component of OAM, $M_{ji}$ is the angular momentum flux density, and $dS_j$ is a surface element whose normal is along $\hat{\mathbf{J}}$. This approach was adopted by Porras in [33] to calculate the tOAM contained in a pulse by time-integrating the tOAM flux through a fixed surface.

The problems in [33] stem from using the flux approach to evaluate tOAM embedded in a transient pulse envelope, and originate in its Eqs. (21) and (23). The first of these equations is reproduced here using that paper's units and notation: $J_y = (\varepsilon_0 z/2k_0)\int |A|^2 \partial_x \Phi d\vec{x}_\perp dt' - (\varepsilon_0/2)\int |A|^2 x\, d\vec{x}_\perp dt'$, where $\Phi = \arg(A)$. This equation, attempting to integrate the tOAM flux at a surface located at z, is clearly in error, mathematically and physically: the first term incorrectly assumes that all values of $z$ are associated with the identical expectation value of transverse linear momentum, $(\varepsilon_0/2k_0)\int |A|^2 \partial_x \Phi d\vec{x}_\perp dt'$, whereas $z$ should be a lever arm weighted *inside* the integral by the linear momentum density $(\varepsilon_0/2k_0)\,|A|^2 \partial_x \Phi$. The same mistake is repeated in [34]. While the flux approach of Barnett is appropriate for calculating the longitudinal OAM of a composite object, it is clearly inappropriate for tOAM. Appendix C presents a simple model clearly illustrating this point.

Correctly leaving $z$ inside the integrals of Eqs. (21) and (23) of [33], applying the author's own transformation $t' = t - z/c$ (using the units and notation of [33]), and normalizing by the number of photons ($U/ck_0$, where $U$ is the pulse energy) gives

$$J_y^{\prime(e)} = \frac{\varepsilon_0 c^2}{2U}(t - t'_m)\int |A|^2 \partial_x \Phi\, d\vec{x}_\perp dt' - \frac{\varepsilon_0 c k_0}{2U}\int |A|^2 x\, d\vec{x}_\perp dt' = 0 \tag{11}$$

$$\text{so that}\quad J_y^{\prime(i)} = \frac{\varepsilon_0 c^2}{2U}\int (t - t')|A|^2 \partial_x \Phi d\vec{x}_\perp dt' - \frac{\varepsilon_0 c k_0}{2U}\int |A|^2 x\, d\vec{x}_\perp dt' = \frac{\alpha}{2}\ , \tag{12}$$

where

$$A = \left(\frac{t'}{w_{0t}} + i\frac{x}{w_{0x}}\right)\exp\left(-\frac{t'^2}{w_{0t}^2} - \frac{x^2}{w_{0x}^2} - \frac{y^2}{w_{0y}^2}\right) \tag{13}$$

and $w_{0t} = w_{0\xi}/c$. Above, $J_y^{\prime(i)}$ and $J_y^{\prime(e)}$ are the intrinsic and extrinsic tOAM per photon. These corrected results are in agreement with our theory [30].

## Appendix C: Simple tOAM model for a composite system

Consider the simple composite system of two point objects shown in Fig. 4. Here $\mathbf{v}_c$ is the velocity of the centre of mass (energy) and $\pm\mathbf{p}$ are the perpendicular momenta of the objects. First, we evaluate the total tOAM with respect to $(x, z) = (0,0)$ by summing contributions *at a common instant of time* when the centre of mass is at $z = l$. This gives $L_y = (l + a)p + (l - a)(-p) = 2ap$, which corresponds to the intrinsic tOAM of the system. This correct result is the particle analogue of the "common instant of time" recipe of Eq. (1). Alternatively, we can sum the OAM contributions as the composite object passes through the plane at $z = l$, which is the precise analogue of what Porras has done in Refs. [33,34,39]. This gives $L_y = lp + l(-p) = 0$, where the two terms are separated in time by $2a/v_c$. This result of zero total OAM occurs for the same reason that Porras obtains it for STOVs: it is an *artifact of adding OAM contributions from different times.* This toy model clearly shows why the flux integration approach of Barnett [38]—in the manner employed by Porras [33,34,39]—gives incorrect results for STOVs.

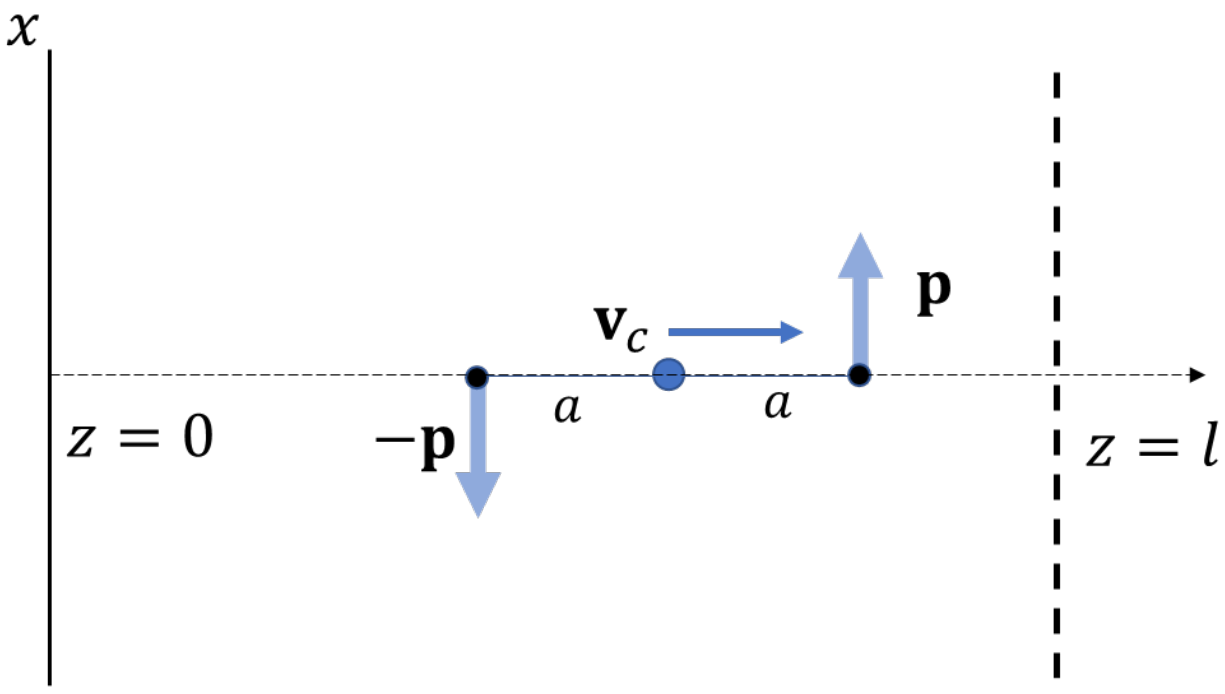

**Fig. 4.** Simple composite tOAM system of two point objects.

**Funding.** Air Force Office of Scientific Research (FA9550-21-1-0405); US Department of Energy (DE-SC0024398 and DE-SC0024406).

**Acknowledgments.** The authors thank Manh Le and Andrew Goffin for useful discussions.

**Disclosures.** The authors declare no conflicts of interest.

**Data availability.** Data underlying the results may be obtained from the authors upon reasonable request.